\documentclass[conference,letterpaper]{IEEEtran}

\usepackage[utf8]{inputenc} 
\usepackage[T1]{fontenc}
\usepackage{url}
\usepackage{ifthen}
\usepackage{cite}
\usepackage[cmex10]{amsmath} % Use the [cmex10] option to ensure complicance
\usepackage{url}
\usepackage{cite}
\usepackage{enumitem}

\usepackage[hidelinks]{hyperref}%
\usepackage[cmex10]{amsmath} 
\usepackage{bbm}
\usepackage[english]{babel}
\usepackage{amsmath,amssymb,amsfonts}
\usepackage{dsfont}
\usepackage{graphicx}
\usepackage{textcomp}
\usepackage{enumerate}
\usepackage{xcolor}

\usepackage{csquotes}
\usepackage{chngcntr}
\usepackage{apptools}
\usepackage{mathtools}
\usepackage{tikz}
\usepackage{caption} 
\usepackage{cleveref}
\usepackage{upgreek}
\usetikzlibrary{mindmap,backgrounds,arrows.meta}
\usepackage{algorithm}
\usepackage{algpseudocode}
\usepackage{enumitem}
\newcommand{\E}{\mathcal{E}}

\newcommand{\Z}{\mathcal{Z}}

\newcommand{\V}{\mathcal{V}}

\makeatletter
\newcounter{labelcnt}
\renewcommand{\thelabelcnt}{(\alph{labelcnt})}
\newcommand{\setlabel}[1]{%
	\refstepcounter{labelcnt}\ltx@label{lbl:#1}%
	{\text{\upshape\thelabelcnt}}%
}

\makeatother
\usepackage[sort&compress,square,numbers]{natbib}

\usepackage{wrapfig} % Allows wrapping text around tables and figures

\usepackage[utf8]{inputenc}

\providecommand{\floor}[1]{\left\lfloor{#1}\right\rfloor}

\usepackage{tikz}
\usetikzlibrary{shapes,arrows}
\usepackage{pgfplots}
\pgfplotsset{compat=1.16}
\usetikzlibrary{shapes.geometric}
\usepackage{xcolor}

\usepackage{tikzscale}

\usepackage{lipsum}
\usepgfplotslibrary{fillbetween}
\usetikzlibrary{intersections}
\usetikzlibrary{patterns}
\pgfdeclarelayer{ft}
\pgfsetlayers{ft,main}
\usepackage[absolute,overlay]{textpos}
\usepackage{graphicx}
\usepackage{caption}
\usepackage{times}
\usepackage{float}
\usepackage{amsmath}
\usepackage{amstext}
\usepackage{amssymb,bm, bbm}
\usepackage{latexsym}
\usepackage{color}
\usepackage{graphicx}
\usepackage{amsmath}
\usepackage{amsthm}
\usepackage{graphicx}
\usepackage{caption}
\usepackage{lipsum}% dummy text
\usepackage{dblfloatfix}
\usepackage{mathrsfs}
\usepackage{cite}
\usepackage{tikz}
\usepackage{pgfplots}
\pgfplotsset{compat=newest}
\usepackage{enumitem}
\usepackage{makecell}
\restylefloat{table}

\allowdisplaybreaks

\theoremstyle{mystyle}
\newtheorem{theorem}{Theorem}%[section]
\theoremstyle{mystyle}
\newtheorem{lemma}{Lemma}%[section]
\newtheorem{remark}{Remark}

\theoremstyle{mystyle}
\newtheorem{prop}{Property}%[section]
\theoremstyle{mystyle}
\theoremstyle{mystyle}
\newtheorem{definition}{Definition}%[section]
\theoremstyle{remark}
\theoremstyle{mystyle}
\theoremstyle{mystyle}
\newtheorem{example}{Example}%[section]
\theoremstyle{mystyle}
\theoremstyle{discussion}
\theoremstyle{mystyle}
\theoremstyle{mystyle}
\theoremstyle{mystyle}

 \usepackage{enumitem}

\usepackage{tabularx}
\usepackage{array}
\usepackage{wrapfig} % Allows wrapping text around tables and figures
\usepackage{multicol}
\usepackage{tabularx}
\usepackage{blindtext}
\usepackage{microtype}
\usepackage{tabu}
\usepackage[utf8]{inputenc}

\begin{document}

\title{On the Secrecy Capacity of 1-2-1 Atomic Networks} 

% \IEEEoverridecommandlockouts

\author{
\IEEEauthorblockN{Mohammad Milanian, Minoh Jeong, Martina Cardone}
University of Minnesota, Minneapolis, \!MN 55455, \!USA,
\!Email: \{milanian, jeong316, mcardone\}@umn.edu
% \\
% \thanks{This research was supported in part by the U.S. National Science Foundation under Grant CCF-2045237.} 
}
\maketitle

\begin{abstract}
We consider the problem of secure communication over a noiseless 1-2-1 network, an abstract model introduced to capture the directivity characteristic of mmWave communications.
We focus on structured networks, which we refer to as 1-2-1 atomic networks. Broadly speaking, these are characterized by a source, a destination, and three layers of intermediate nodes with sparse connections.
The goal is for the source to securely communicate to the destination in the presence of an eavesdropper with unbounded computation capabilities, but limited network presence. We derive novel upper and lower bounds on the secrecy capacity of 1-2-1 atomic networks. These bounds are shown to be tighter than existing bounds in some regimes. Moreover, in such regimes, the bounds match and hence, they characterize the secrecy capacity of 1-2-1 atomic networks.
\end{abstract}

\section{Introduction}
Millimeter Wave (mmWave) communication has shown a great potential in overcoming the spectrum scarcity and enabling multi-gigabit services~\cite{Wang2018},~\cite{Uwaechia2020}.
In~\cite{ezzeldin2020gaussian}, the authors introduced the so-called  {\em{1-2-1 network}} model with the goal to study the information theoretic capacity of mmWave networks.
This model abstracts away the physical layer component, while capturing the fundamental directivity characteristic of mmWave communication.
In the 1-2-1 network model, in order to activate the communication link from node $a$ to node $b$, these two nodes need to perform beamforming in a way that their beams face each other (hence, the term 1-2-1).

In this work, similarly to~\cite{agarwal2018secure} and~\cite{EzzeldinSecurity}, we study secure communication over 1-2-1 networks. 
We consider 1-2-1 networks with lossless communication links of unitary capacity, where a source wishes to securely communicate with a destination in the presence of an eavesdropper with limited network coverage, but unbounded computation capabilities (e.g., quantum computer). In particular, the adversary can eavesdrop at most $K$ edges of their choice. This assumption on the limited network presence is reasonable in mmWave networks since the adversary has to be physically present on a link to eavesdrop the high-directional communication over it. We assume that the source and destination have stronger beamforming capabilities than the intermediate nodes, i.e., they can transmit to (the source) and receive from (the destination) at most $M$ nodes, whereas the intermediate nodes have only a single transmit beam and a single receive beam.
We focus on a class of 1-2-1 networks, which we refer to as 1-2-1 {\em atomic} networks, an example of which is shown in Figure~\ref{fig:atomic}.
Broadly speaking, these are layered networks (no communication links exist between nodes within the same layer). Our main contribution consists of: (1) deriving novel lower and upper bounds on the secrecy capacity of 1-2-1 atomic networks; and (2) providing conditions under which these match, hence characterizing the secrecy capacity in these regimes. The lower bound is obtained through the design of a transmission scheme that suitably leverages the network {\em multipath} to establish keys and transmit messages (encoded with the keys) between the source and the destination. A surprising result of our work is that, given the same network topology, the source may need to change the transmission strategy as $K$ increases.

We study secure communication over {1-2-1} networks as~\cite{agarwal2018secure} and~\cite{EzzeldinSecurity}.
However, our work has key distinguishing features from~\cite{agarwal2018secure} and~\cite{EzzeldinSecurity}. In~\cite{EzzeldinSecurity}, the 1-2-1 model analyzed is different from the 1-2-1 atomic network considered in this paper. In~\cite{agarwal2018secure}, although the derived lower and upper bounds can be applied to 1-2-1 atomic networks, they are not tight in general.

In order to ensure secrecy, we here leverage the following two aspects: 
(1) directivity; and
(2) multipath.
{\em Directivity} has been investigated for ensuring security in MIMO beamforming~\cite{Mukherjee2011},~\cite{safaka2016creating}. The main idea of these works is to create beams that are narrow enough to significantly weaken the channel of the adversary. However, these works focus on guaranteeing secrecy over channels and not over networks, which is our goal in this paper.
{\em Multipath} has also been leveraged for security in noiseless networks in the context of secure network coding~\cite{cai2002secure,khaleghi2009subspace,jaggi2007resilient,jaggi2005polynomial,safaka2016creating,agarwal2017secure,cui2012secure}. The literature on (linear) secure network coding is rich, with a few examples given by~\cite{cai2007security,ngai2011network,wei1991generalized,koetter2004network,gadouleau2011graph,jaggi2007resilient,ho2004byzantine,papadopoulos2015lp,czap2014triangle}. However, these works consider networks in which nodes can simultaneously communicate to all the connected nodes. Differently, in this work we have a directivity (1-2-1) constraint imposed by mmWave communication, which allows each node only to communicate with a limited number of connected nodes.

\noindent{\bf{Paper organization.}} In Section~\ref{sec:Model}, we present the 1-2-1 atomic network model and we review a few existing results on the secrecy capacity of 1-2-1 networks.
In Section~\ref{sec:MainRes}, we describe our main results, i.e., we derive novel upper and lower bounds on the secrecy capacity and identify regimes in which they are tight. 
In Section~\ref{sec:ProofMainRes}, we provide the proof of our main results. 
We delegate some of the proofs to the appendix.

\noindent{\bf{Notation.}}
For any $m \in \mathbb{N}$, we define ${[m] := \{1,2, \ldots,m\}}$; 
$[a:b]$ is the set of integers from $a$ to $b \geq a$;
$[x]^+ = \max \{0,x\}$.
For a set $\mathcal{X}$, $|\mathcal{X}|$ denotes its cardinality;
$\varnothing$ is the empty set. For a real number $x \in \mathbb{R}$, we denote its floor by $\floor{x}$. We use $X_{\mathcal{S}}$ to denote $(X_{j}: j \in \mathcal{S})$.

\section{System Model and Known Results}
\label{sec:Model}
We consider a 1-2-1 network~\cite{ezzeldin2020gaussian}, where a source $\mathsf{S}$ wishes to securely communicate with a destination $\mathsf{D}$.
The network is modeled by a directed acyclic graph $G = (\mathcal{V},\mathcal{E})$, where $\mathcal{V}$ is the set of vertices such that $(\mathsf{S},\mathsf{D}) \in \mathcal{V}$ and $\mathcal{E}$ is the set of edges.
All edges are lossless and have a fixed finite capacity, which we assume to be {\em unitary}, without loss of generality.

\begin{figure}[t]
\centering
\resizebox{0.4\textwidth}{!}{%
     
 \tikzstyle{block} = [circle , draw,  
    text width=1em, text centered]
    \tikzstyle{bblock} = [circle , draw, fill=blue!20, 
    text width=.2em, text centered]
\tikzstyle{line} = [draw, -latex']
 
\tikzstyle{root} = [,]
\begin{tikzpicture}
    % Place nodes
\tikzstyle{every node}=[font=\small]
\node [fill=blue!30, block] (S) {S};
\node[, block, right of=S, xshift=1cm, yshift=2.5cm] (a) {$a$};
\node[, block, right of=S, xshift=1cm, yshift=1.5cm] (b) {$b$};
\node[, block, right of=S, xshift=1cm, yshift=.5cm] (c) {$c$};
\node[, block, right of=S, xshift=1cm, yshift=-.5cm] (d) {$d$};
\node[, block, right of=S, xshift=1cm, yshift=-1.5cm] (e) {$e$};
\node[, block, right of=S, xshift=1cm, yshift=-2.5cm] (d') {$f$};
\node[, block, right of=S, xshift=1cm, yshift=-3.5cm] (e') {$g$};
\node[fill=green!80, block, right of=b, xshift=.8cm, yshift=-.3cm] (f) {$v_{1}$};
\node[fill=green!80, block, right of=d, xshift=.8cm, yshift=-.1cm] (g) {$v_{2}$};

\node[, block, right of=S, xshift=4.6cm, yshift=2.5cm] (h) {$h$};
\node[, block, right of=S, xshift=4.6cm, yshift=1.5cm] (i) {$i$};
\node[, block, right of=S, xshift=4.6cm, yshift=.5cm] (j) {$j$};
\node[, block, right of=S, xshift=4.6cm, yshift=-.5cm] (k) {$k$};
\node[, block, right of=S, xshift=4.6cm, yshift=-1.5cm] (l) {$l$};
\node[, block, right of=S, xshift=4.6cm, yshift=-2.5cm] (k') {$m$};
\node[, block, right of=S, xshift=4.6cm, yshift=-3.5cm] (l') {$n$};
\node[fill=green!80, block, right of=d', xshift=.8cm, yshift=.3cm] (g') {$v_{3}$};
\node [fill=blue!30, block, right of=S, xshift=6.6cm] (D) {$\mathsf{D}$};

\draw[<-,shorten <=0pt,shorten >=0pt, >=stealth', semithick] (a) -- (S)node[midway,above] {};
\draw[<-,shorten <=0pt,shorten >=0pt, >=stealth', semithick] (b) -- (S)node[midway,above] {};
\draw[<-,shorten <=0pt,shorten >=0pt, >=stealth', semithick] (c) -- (S)node[midway,above] {};
\draw[<-,shorten <=0pt,shorten >=0pt, >=stealth', semithick] (d) -- (S)node[midway,above] {};
\draw[<-,shorten <=0pt,shorten >=0pt, >=stealth', semithick] (e) -- (S)node[midway,above] {};
\draw[<-,shorten <=0pt,shorten >=0pt, >=stealth', semithick] (d') -- (S)node[midway,above] {};
\draw[, thick, <-,shorten <=0pt,shorten >=0pt, >=stealth', semithick] (e') -- (S)node[midway,above] {};
\draw[<-,shorten <=0pt,shorten >=0pt, >=stealth', semithick] (g') --(d')node[midway,above] {};
\draw[<-,shorten <=0pt,shorten >=0pt, >=stealth', semithick] (g') --(e')node[midway,above] {};

\draw[<-,shorten <=0pt,shorten >=0pt, >=stealth', semithick] (f) -- (a)node[midway,above] {};
\draw[<-,shorten <=0pt,shorten >=0pt, >=stealth', semithick] (f) -- (b)node[midway,above] {};
\draw[, thick, <-,shorten <=0pt,shorten >=0pt, >=stealth', semithick] (f) -- (c)node[midway,above] {};
\draw[<-,shorten <=0pt,shorten >=0pt, >=stealth', semithick] (g) -- (d)node[midway,above] {};  
\draw[, thick, <-,shorten <=0pt,shorten >=0pt, >=stealth', semithick] (g) -- (e)node[midway,above] {};
\draw[<-,shorten <=0pt,shorten >=0pt, >=stealth', semithick] (k') --(g')node[midway,above] {};
\draw[<-,shorten <=0pt,shorten >=0pt, >=stealth', semithick] (l') --(g')node[midway,above] {};
\draw[<-,shorten <=0pt,shorten >=0pt, >=stealth', semithick] (h) -- (f)node[midway,above] {};
\draw[<-,shorten <=0pt,shorten >=0pt, >=stealth', semithick] (i) -- (f)node[midway,above] {};
\draw[<-,shorten <=0pt,shorten >=0pt, >=stealth', semithick] (j) -- (f)node[midway,above] {};
\draw[<-,shorten <=0pt,shorten >=0pt, >=stealth', semithick] (k) -- (g)node[midway,above] {};
\draw[<-,shorten <=0pt,shorten >=0pt, >=stealth', semithick] (l) -- (g)node[midway,above] {};
\draw[<-,shorten <=0pt,shorten >=0pt, >=stealth', semithick] (D) -- (h)node[midway,above] {};
\draw[<-,shorten <=0pt,shorten >=0pt, >=stealth', semithick] (D) -- (i)node[midway,above] {};
\draw[<-,shorten <=0pt,shorten >=0pt, >=stealth', semithick] (D) -- (j)node[midway,above] {};
\draw[, thick, <-,shorten <=0pt,shorten >=0pt, >=stealth', semithick] (D) -- (k)node[midway,above] {};
\draw[<-,shorten <=0pt,shorten >=0pt, >=stealth', semithick] (D) -- (l)node[midway,above] {};
\draw[, thick, <-,shorten <=0pt,shorten >=0pt, >=stealth', semithick] (D) -- (k')node[midway,above] {};
\draw[<-,shorten <=0pt,shorten >=0pt, >=stealth', semithick] (D) -- (l')node[midway,above] {};
\end{tikzpicture}   
}
\caption{A 1-2-1 atomic network with $H_e=7$ and $H_v=3$. The $H_e=7$ edge-disjoint paths are referred to as $p_i, i \in [7]$ from top to bottom, e.g., $p_4 = \mathsf{S} \to d \to v_2 \to k \to \mathsf{D}$.}
\label{fig:atomic}  
 \end{figure}
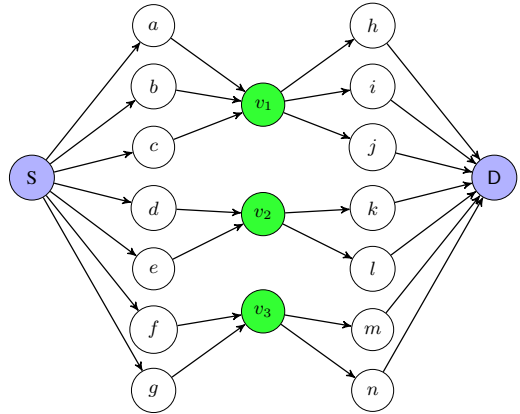

An edge $e\in\mathcal{E}$ is activated according to the 1-2-1 constraint. Under this constraint, each intermediate node (i.e., each $v \in \mathcal{V}$ except for $\mathsf{S}$ and $\mathsf{D}$) can simultaneously receive and transmit, but at each point in time it can receive from at most one incoming edge and transmit through at most one outgoing edge. For instance, with reference to Figure~\ref{fig:atomic}, at each point in time, node $v_1$ can receive from at most one among nodes $a, b ,c$ and transmit to at most one among nodes $h,i,j$.
Differently, the source $\mathsf{S}$ (respectively, the destination $\mathsf{D}$) can transmit to (respectively, receive from) at most $M \geq 1$ nodes.

We denote by $H_e$ and $H_v$ the maximum numbers of edge-disjoint and vertex-disjoint paths from $\mathsf{S}$ to $\mathsf{D}$, respectively.
Among arbitrary 1-2-1 networks, our focus in this paper is on structured networks, which we refer to as 1-2-1 {\em atomic} networks, as formally defined next.
\begin{definition}\label{atomic}
A 1-2-1 atomic network is one for which its underlying $G = (\mathcal{V},\mathcal{E})$
can be partitioned into $H_v$ atomic subgraphs\footnote{Even though we refer to the $G_i$'s as a partition of $G$, we assume that $\mathsf{S}$ and $\mathsf{D}$ are common to all the subgraphs.} $G_i =(\mathcal{V}_{i},\mathcal{E}_{i})$ with $\V_{i} \subseteq \V$ and $\E_{i} \subseteq \E$ for all $i\in[H_v]$, such that the three following conditions hold:
\begin{itemize}
    \item The maximum number of vertex disjoint paths (respectively, edge disjoint paths) from $\mathsf{S}$ to $\mathsf{D}$ in each $G_{i}$ is equal to one (respectively, $h_{i}$);
    % \item Each $G_i$ has one vertex disjoint path from $\mathsf{S}$ to $\mathsf{D}$;
    \item All the $h_i$ edge disjoint paths in each $G_{i}$ from $\mathsf{S}$ to $\mathsf{D}$ only share one (intermediate) node, referred to as \textit{atom} $v_{i}$;
    \item Any two $G_{i}$ and $G_{j}$ share no nodes other than $\mathsf{S}$ and $\mathsf{D}$, for all $i, j \in [H_{v}], i \neq j$.
\end{itemize}
\end{definition}
Throughout the paper, we will represent a 1-2-1 atomic network by a vector $\mathbf{h}=[h_1,\ldots,h_{H_v}]$, where $h_i$ is defined in Definition~\ref{atomic}.
Figure~\ref{fig:atomic} provides an example of a 1-2-1 atomic network with $H_v=3$ and $\mathbf{h}=[3,2,2]$.

The communication from $\mathsf{S}$ to $\mathsf{D}$ over a 1-2-1 atomic network takes place in the presence of an {\em external passive adversary} who can eavesdrop any $K$ edges of their choice (unknown to all the other nodes in the network).
If the adversary eavesdrops edges in $\Z \subseteq \mathcal{E}, |\Z|=K$, we require that the communication remains secure from the adversary in the following sense,
    \begin{align}\label{eq:sec}
        I(W;T^{[n]}_{\mathcal{Z}}) \leq \varepsilon, ~~\forall \mathcal{Z}\subseteq \mathcal{E}, ~|\mathcal{Z}| = K,
    \end{align}
where $W$ is the message with entropy rate $R$ that $\mathsf{S}$ wishes to transmit to $\mathsf{D}$ and $T^{[n]}_{\mathcal{Z}} = \left \{ T_e^{[n]}, e \in \mathcal{Z}\right \}$ denotes the packets transmitted over $e \in \mathcal{Z}$ in $n$ network uses.

In this paper, we seek to characterize the secrecy capacity $\mathsf{C}_s$ for 1-2-1 atomic networks. This is defined as the maximum rate at which $\mathsf{S}$ can communicate to $\mathsf{D}$ with zero error, while satisfying the 1-2-1 constraints of the network and the security constraint in~\eqref{eq:sec}.
To the best of our knowledge, the tightest bounds on $\mathsf{C}_s$ are given by~\cite{agarwal2018secure},
\begin{equation}
\label{eq:Bounds}
\min \left \{ M,H_v\right \}\frac{H_v-K}{H_v} \leq \mathsf{C}_s \leq \min \left \{ M,H_e \right \} \frac{H_e - K}{H_e}.
\end{equation}
It is worth noting that the lower and the upper bounds in~\eqref{eq:Bounds} match when $H_e = H_v$. However, when $H_v < H_e$ we will show that, in general,
neither the lower bound nor the upper bound in~\eqref{eq:Bounds} are tight for a 1-2-1 atomic network. 
For instance, consider the 1-2-1 atomic network in Figure~\ref{fig:atomic} for which $H_e = 7$ and $H_v=3$. Assume that $K=1$ and $M=3$. From the bounds in~\eqref{eq:Bounds}, we obtain that $2 \leq \mathsf{C}_s \leq 18/7$. However, in this paper we will prove that $\mathsf{C}_s=5/2$.

%%%%%%%%%%%%%%%%%%%%%%%%%%%%%%% Fig.

\section{Main Results}
\label{sec:MainRes}
In this section, we present the main results of our work. 
We start by providing a lower bound on the secrecy capacity $\mathsf{C}_s$ of a 1-2-1 atomic network.
\begin{theorem}
\label{thm:LB}
Define $\widehat{M}:= \min (M,H_{v})$ and,
for all $c\in[0:\widehat{M}-1]$, let
\begin{align}
\label{eq:Achievable}
    R(c)
    & = \left [ \frac{\sum_{\eta=0}^{c}(\widehat{M}-\eta)\mathsf{P}_{\widehat{M}-\eta} - K}{\sum_{\eta=0}^{c}\mathsf{P}_{\widehat{M}-\eta}} \right ]^+,
\end{align}
where $\mathsf{P}_{\widehat{M}-\eta}:=|\mathcal{P}_{\widehat{M}-\eta}|$ with $\mathcal{P}_{\widehat{M}-\eta}, \eta \in [0:c]$ being a set\footnote{Algorithm~\ref{alg:algo1} will provide the detailed construction of $\mathcal{P}_{\widehat{M}-\eta}$.} with 
% the maximum number of 
groups of $\widehat{M}-\eta$ vertex-disjoint paths of $G\setminus \bigcup_{i=0}^{\eta-1} \mathcal{P}_{\widehat{M}-i}$. Each path can appear in at most one group.

Then, for a 1-2-1 atomic network with $K$ eavesdropped edges, it holds that $\mathsf{C}_s \geq R(c^\star)$ where $c^\star = \arg \max_{c\in[0:\widehat{M}-1]} R(c)$.
\end{theorem}
\begin{IEEEproof}
See Section~\ref{sec:ProofThmLB}.
\end{IEEEproof}
\begin{example}
\label{Ex:cstar}
Consider the 1-2-1 atomic network in Figure~\ref{fig:atomic} with $M=3$. Thus, $\widehat{M}=3$. Assuming $K=1$, we have that:

\noindent $1) \ \eta=0:$ $\mathcal{P}_{3} \!=\! \left \{\{p_1,p_4,p_6\},\{p_2,p_5,p_7\} \right \}$ and hence, $\mathsf{P}_{3} \!=\! 2$; 

\noindent $2) \ \eta=1:$ $\mathcal{P}_{2} = \varnothing$ and hence, $\mathsf{P}_{2} = 0$;

\noindent $3) \ \eta=2:$ $\mathcal{P}_{1} = \left \{ p_3\right \}$ and hence, $\mathsf{P}_{1} = 1$.

By using the above inside~\eqref{eq:Achievable}, we obtain $R(0) = 5/2$, $R(1)=5/2$, and $R(2)=2$. Thus, $c^\star = 0$ and $\mathsf{C}_s \geq 5/2$.
\end{example}

\begin{remark}
The value of $c^\star$ in Theorem~\ref{thm:LB} for a specific 1-2-1 atomic network depends on $K$. For instance, assume the same setting as in Example~\ref{Ex:cstar}, but with $K=5$. Then, we would obtain $R(0) = 1/2$, $R(1)=1/2$, and $R(2)=2/3$. Thus, $c^\star = 2$, which is different from $c^\star = 0$ when $K=1$.
\end{remark}

We observe that the lower bound on $\mathsf{C}_s$ in Theorem~\ref{thm:LB} depends on $\mathsf{P}_{\widehat{M}-\eta}, \eta \in [0:c]$. Thus, a natural question arises: Is it possible to find an expression to compute $\mathsf{P}_{\widehat{M}-\eta}, \eta \in [0:c]$? The next lemma will be helpful to provide an answer to this question.
\begin{lemma}\label{lem:t}  
Consider a 1-2-1 atomic network with underlying graph $G$. Assume, without loss of generality, that $h_1 \geq h_2 \geq \ldots \geq h_{H_v}$.
Then, there are at least $\mathsf{P}_{\ell}(h_{[H_v]})$ groups of $\ell$ vertex-disjoint paths in $G$, where
%Let $\mathsf{P}_{i}(h_{[H_v]}):=|\mathcal{P}_{i}(h_{[H_v]})|,~i \in [H_v]$,
%with $\mathcal{P}_{i}(h_{[H_v]}), i \in [H_v]$, being a set with groups of $i$ vertex-disjoint paths of $G$.
%{\color{cyan}
%Let $\mathsf{P}_{i}(h_{[H_v]}):=|\mathcal{P}_{i}(h_{[H_v]})|,~i \in [H_v]$,
% with $\mathcal{P}_{i}(h_{[H_v]}), i \in [H_v]$, 
%where $\mathcal{P}_{i}(h_{[H_v]}) := \mathcal{P}_{i}$ is defined in Theorem~\ref{thm:LB} with representation $h_{[H_v]}$ of $G\setminus \bigcup_{j=i+1}^{H_v} \mathcal{P}_{j}$. Assume, without loss of generality, that $h_1 \geq h_2 \geq \ldots \geq h_{H_v}$.}
% being the set with groups of $i$ vertex-disjoint paths of $G$.
% Each path can appear in at most one group.
%Then, it holds that
\begin{subequations}
\begin{align}
\mathsf{P}_1 \left (h_{[H_v]} \right ) = \sum_{i=1}^{H_v} h_{i},
\end{align}
and, for all $\ell \in [2:H_v]$, it holds that
\begin{align}
\mathsf{P}_{\ell} \left (h_{[H_v]} \right ) = 
\left \{
\begin{array}{ll}
\mathsf{P}_{\ell-1} \left (h_{[2:H_v]} \right ) & \text{if } h_1 \geq \mathsf{P}_{\ell-1} \left (h_{[2:H_v]} \right ),
\\
\left \lfloor\frac{\sum_{i=1}^{H_v} h_{i}}{ \ell }\right\rfloor & \text{otherwise.}
\end{array}
\right .
\end{align}
\end{subequations}
\end{lemma}
\begin{IEEEproof}
See Appendix~\ref{app:proof_t}. 
\end{IEEEproof}
We note that Lemma~\ref{lem:t} can be applied $\widehat{M}$ times to compute $\mathsf{P}_1,\mathsf{P}_2, \ldots, \mathsf{P}_{\widehat{M}}$ in Theorem~\ref{thm:LB}. In particular, to compute $\mathsf{P}_{\widehat{M}}$ Lemma~\ref{lem:t} is applied over $G$; then, to compute $\mathsf{P}_{\widehat{M}-1}$ Lemma~\ref{lem:t} is applied over $G \setminus \mathcal{P}_{\widehat{M}}$ and so on, until $\mathsf{P}_{1}$ that can be computed by applying Lemma~\ref{lem:t} over $G\setminus \bigcup_{i=0}^{\widehat{M}-2} \mathcal{P}_{\widehat{M}-i}$.

We now focus on deriving an upper bound on the secrecy capacity $\mathsf{C}_s$ of a 1-2-1 atomic network.
\begin{theorem}
\label{thm:UB}
For a 1-2-1 atomic network with $K$ eavesdropped edges, it holds that
\begin{subequations}
\label{eq:UBThm}
\begin{align}
    \mathsf{C}_s
    & \leq \sum_{i=1}^{H_v} \left( 1 - \frac{K_i}{h_i}\right),
\end{align}
where, without loss of generality, it is assumed that $h_1 \geq h_2 \geq \ldots \geq h_{H_v}$, and where
\begin{align}
    K_i
    & = \min \left\{  h_i, K - \sum_{j=i+1}^{H_v} K_j \right\},~ i\in[H_v].
\end{align}
\end{subequations}
\end{theorem}
\begin{IEEEproof}
See Section~\ref{sec:ProofThmUB}.
\end{IEEEproof}
We now leverage the results in Theorem~\ref{thm:LB} and Theorem~\ref{thm:UB}
to prove the following secrecy capacity result.
\begin{theorem}
\label{thm:SecCap}
For a 1-2-1 atomic network with $K$ eavesdropped edges and $M\geq H_v$, the derived achievability and converse bounds match and hence, under this condition, the secrecy capacity of the considered network is given by~\eqref{eq:UBThm}.
\end{theorem}
\begin{IEEEproof}
See Section~\ref{sec:ProofThmCap}.
\end{IEEEproof}
Theorem~\ref{thm:SecCap} provides a new secrecy capacity result for 1-2-1 networks. 
In particular, when $M \geq H_v$ our lower and outer bounds in Theorem~\ref{thm:LB} and Theorem~\ref{thm:UB} are tighter than those in~\eqref{eq:Bounds}.
However, we next show that when $M<H_v$ the bounds in~\eqref{eq:Bounds} may be tighter. This implies that our bounds in Theorem~\ref{thm:LB} and Theorem~\ref{thm:UB} can be further improved, and this is indeed object of current investigation.
\begin{example}
Consider a 1-2-1 atomic network with $\mathbf{h}=[2,1,1,1]$, $M=3$ and $K=1$.
From Theorem~\ref{thm:LB}, we obtain $\mathsf{C}_s \geq 2$, whereas from~\eqref{eq:Bounds} we have that  $\mathsf{C}_s \geq 9/4$. Thus, our achievable bound in looser than the existing one from~\cite{agarwal2018secure}.
\end{example}
\begin{example}
Consider a 1-2-1 atomic network with $\mathbf{h}=[4,3,2]$, $M=2$, and $K=5$. 
From Theorem~\ref{thm:UB}, we obtain $\mathsf{C}_s \leq 1$, whereas from~\eqref{eq:Bounds} we have that  $\mathsf{C}_s \leq 8/9$. Thus, our converse bound in looser than the existing one from~\cite{agarwal2018secure}.
\end{example}

\section{Proof of Main Results}
\label{sec:ProofMainRes}

\subsection{Proof of Theorem~\ref{thm:LB}}
\label{sec:ProofThmLB}
We here propose a secure transmission scheme for a 1-2-1 atomic network with $K$ eavesdropped edges and we prove that it achieves the secrecy rate in Theorem~\ref{thm:LB}. 
The scheme consists of four phases, which are next described.
The key generation and encoding phases are the same as those in~\cite{agarwal2018secure}.

\noindent{\bf 1) Key generation.}
We generate $K$ uniform random packets, denoted by $X_{[K]}$ and create $H_e$ linear combinations of them by pre-multiplying $X_{[K]}$ by a maximum distance separable (MDS) code matrix $V$ of size $H_e \times K$, i.e., 
\begin{align}\label{eq:rand_pck}
    f(X) = VX_{[K]}.
\end{align}
In what follows, we will refer to each row of $f(X)$ in~\eqref{eq:rand_pck} as a {\em key}. 
Note that any $K$ rows of $f(X)$ are linearly independent.
%{\color{red}Note that the $H_e$ rows of $f(X)$ are mutually independent.}

\noindent{\bf 2) Encoding.}
We take $H_e - K$ message packets $W_j, j \in [H_e - K]$ and we encode them using $f(X)$ in~\eqref{eq:rand_pck}. In particular, this encoding operation is as follows,
\begin{align}\label{eq:encoder}
    T_{i}
    & = \begin{cases}
        f_{i}(X) &\quad i \in [K], \\
        f_{i}(X)+ W_{i-K} &\quad i \in [K+1:  H_e],
     \end{cases}
\end{align}
where $f_i(X), i \in [H_e]$ is the $i$th row of $f(X)$ in~\eqref{eq:rand_pck}.

\noindent{\bf 3) Transmission.}
The transmission phase consists of $c+1$ rounds of sub-transmissions.
At each round $\eta\in[0:c]$, we first construct a set $\mathcal{P}_{\widehat{M}-\eta}$ of groups of $\widehat{M}-\eta$ vertex-disjoint paths unused from previous rounds. Then, we transmit packets over the paths $p\in\mathcal{P}_{\widehat{M}-\eta}$.
In particular, we construct the set $\mathcal{P}_{\widehat{M}-\eta}, \eta \in [0:c],$ using Algorithm~\ref{alg:algo1}.

\begin{algorithm}
\caption{Construction of $\mathcal{P}_{\widehat{M}-\eta}, \eta \in [0:c]$}\label{alg:algo1}

\begin{algorithmic}[1]
\State Let  $G^{(\eta)} = G\setminus \bigcup_{i=0}^{\eta-1}\mathcal{P}_{\widehat{M}-i}$, which is a graph $G$ with edges that have not been used during rounds $i\in[0:\eta-1]$. 
\State Initialize $\mathcal{P}_{\widehat{M}-\eta} = \varnothing$.
\State Select the $\widehat{M}-\eta$ atomic subgraphs of $G^{(\eta)}$ with maximum number of non-zero edge-disjoint paths.
\State Select one path from each selected atomic subgraph and let $\mathcal{Q}$ denote them. Note that $|\mathcal{Q}| = \widehat{M}-\eta$.
\State Update $\mathcal{P}_{\widehat{M}-\eta} = \mathcal{P}_{\widehat{M}-\eta}\bigcup \mathcal{Q}$ and remove the paths in $\mathcal{Q}$ from $G^{(\eta)}$, i.e., $G^{(\eta)} = G^{(\eta)}\setminus \mathcal{Q}$.
\State Repeat steps 3-5 until step 3 can not run.
\end{algorithmic}
\end{algorithm}

Once $\mathcal{P}_{\widehat{M}-\eta},~\eta\in[0:c]$ are constructed, the source $\mathsf{S}$ starts sending the packets $T_i,~i\in[H_e]$ in~\eqref{eq:encoder} sequentially.
In particular, for $\eta = 0$, $\mathsf{S}$ sends $\widehat{M}$ packets simultaneously (i.e., in one network use) through each of the $|\mathcal{P}_{\widehat{M}}|  =: \mathsf{P}_{\widehat{M}}$ groups of paths in $\mathcal{P}_{\widehat{M}}$. Note that this is possible since: (i) at each point in time, $\mathsf{S}$ can simultaneously transmit to $\widehat{M}$ nodes, and (ii) each group of paths in $\mathcal{P}_{\widehat{M}}$ are vertex-disjoint. Thus, the 1-2-1 constraint is satisfied. At the end of round $\eta = 0$, $\mathsf{S}$ has sent $T_i, i \in [\widehat{M} \mathsf{P}_{\widehat{M}}]$, i.e., round $\eta = 0$ consists of $\mathsf{P}_{\widehat{M}}$ network uses.
After this, round $\eta = 1$ starts and $\mathsf{S}$ sends $\widehat{M}-1$ packets simultaneously (i.e., in one network use) through each of the $|\mathcal{P}_{\widehat{M}-1}|  =: \mathsf{P}_{\widehat{M}-1}$ groups of paths in $\mathcal{P}_{\widehat{M}-1}$.
In particular, round $\eta = 1$ consists of $\mathsf{P}_{\widehat{M}-1}$ network uses.
Then, round $\eta = 2$ will start and so on until $\eta = c$. Each round $\eta \in [0:c]$ consists of $\mathsf{P}_{\widehat{M}-\eta}$ network uses in each of which $\widehat{M}-\eta$ packets are sent by $\mathsf{S}$.
Thus, for each $c\in[0:\widehat{M}-1]$ a total of $\sum_{\eta=0}^{c}(\widehat{M}-\eta)\mathsf{P}_{\widehat{M}-\eta}$ packets are sent by $\mathsf{S}$ in $\sum_{\eta=0}^{c}\mathsf{P}_{\widehat{M}-\eta}$ network uses.

\noindent {\bf 4) Decoding.}
At the destination $\mathsf{D}$, the decoding is done by first finding the $K$ random packets $X_{[K]}$ and then reconstructing the keys $f(X)$. Specifically, since the first $K$ received packets are just keys without messages (see~\eqref{eq:encoder}), the random packets $X_{[K]}$ can be obtained as follows,
\begin{align}\label{eq:decode1}
    X_{[K]}
    & = (V_{[K],[K]})^{-1} T_{[K]} ,
\end{align}
where $V_{[K],[K]}$ is the sub-matrix of $V$ obtained by just retaining the first $K$ rows and 
%the first 
all the $K$ columns of $V$. 
Then, $\mathsf{D}$ can generate the keys $f(X)$ using $X_{[K]}$ in~\eqref{eq:decode1} similar to~\eqref{eq:rand_pck}.
Finally, $\mathsf{D}$ decodes the messages $W_j, j \in [H_e-K]$ as follows,
\begin{align}
\label{eq:decode2}
    \widehat{W}_{i-K}
    & = T_{i} - f_i(X),
\end{align}
where $i \in [K+1:H_e]$.

\noindent
{\bf Security.}
In each network use, the adversary can receive a packet passing through an eavesdropped edge if the eavesdropped edge belongs to the paths used in that particular network use.
Since the $K$ eavesdropped edges can at most be part of $K$ paths, the eavesdropper will receive at most $K$ packets, which are linearly independent thanks to the property of MDS codes (see~\eqref{eq:encoder}).
%symbols, 
%all of which are encoded with independent keys (see~\eqref{eq:encoder}). 
Thus, the scheme securely transmits a total of $\left [ \sum_{\eta=0}^{c}(\widehat{M}-\eta)\mathsf{P}_{\widehat{M}-\eta} -K \right ]^+$ message packets in $\sum_{\eta=0}^{c}\mathsf{P}_{\widehat{M}-\eta}$ network uses.
This leads to $R(c)$ in~\eqref{eq:Achievable}. The proof of Theorem~\ref{thm:LB} is concluded by considering the $c^\star\in[0:\widehat{M}-1]$ for which $R(c^\star)$ is maximum.

\begin{example}\label{ex:scheme}
Consider the 1-2-1 atomic network in Figure~\ref{fig:atomic} with $M=2$. Thus, $\widehat{M}=2$. Assume $K=5$.
Then, the proposed scheme for $c=1$ runs as follows,
\begin{itemize}
    \item[1)] We generate $K=5$ uniform random packets $X_{[5]}$, and extend them to $7$ keys, $f_{i}(X)$, $i\in [7]$ using an MDS code matrix of size $7\times5$.
    \item[2)] We encode $H_e - K = 2$ messages $W_i,~i\in [2]$ as follows,
    \begin{align}\label{eq:encode_example}
        T_{i}= 
        \begin{cases}
            f_{i}(X) &\quad i \in [5], \\
            f_{i}(X)+ W_{i-5} &\quad i \in [6:7].
        \end{cases}
    \end{align}
    \item[3)] From Algorithm~\ref{alg:algo1}, for $c=1$, we obtain 
    \begin{align}
    \mathcal{P}_2 &= \left \{ \left \{ p_1, p_4\right \},\left \{ p_2, p_6\right \},\left \{ p_3, p_5\right \} \right \},
    \\ \mathcal{P}_1 &= \left \{ p_7\right \}.
    \end{align}
    For transmission, we use the network $\mathsf{P}_2 + \mathsf{P}_1 = 3+1=4$ times: (1) each group of paths in $\mathcal{P}_2$ can be used to simultaneously transmit $\widehat{M}=2$ packets, e.g., $p_1$ and $p_4$ can be used to transmit $T_1$ and $T_2$ in the first network use, $p_2$ and $p_6$ can be used to transmit $T_3$ and $T_4$ in the second network use, and $p_3$ and $p_5$ can be used to transmit $T_5$ and $T_6$ in the third network use; and (2) each group of paths in $\mathcal{P}_1$ can be used to simultaneously transmit $\widehat{M}-1=1$ packet, e.g., $p_7$ can be used to transmit $T_7$ in the fourth network use.
    \item[4)] Upon receiving $T_{i}, i \in [7]$, $\mathsf{D}$ recovers $W_1$ and $W_2$ using the property of the MDS code matrix (see~\eqref{eq:decode1} and~\eqref{eq:decode2}).  
\end{itemize}
The adversary can learn at most $5$ packets, which are encoded with independent keys. Thus, the eavesdropper cannot learn anything about $W_1$ and $W_2$. For $c=1$, we hence obtain a secrecy rate $R(1) = \frac{7-5}{4} = \frac{1}{2}$.
\end{example}

\subsection{Proof of Theorem~\ref{thm:UB}}
\label{sec:ProofThmUB}
We let $\mathcal{T}_{\mathcal{S}}^{[n]}$ be the set of packets sent over edges $e \in \mathcal{S}$ in $n$ network uses, i.e., $\mathcal{T}_{\mathcal{S}}^{[n]} = \left \{\mathcal{T}_e^{[n]}, \forall e \in \mathcal{S} \right \}$, and we let $\mathcal{E}^-_{\mathsf{D}}$ be the set of all edges incoming into $\mathsf{D}$. 
We also let $\mathcal{T}_{\mathcal{E}_{i}}^{[n]}$ be the packets sent to atom $v_{i} \in \V_{i}$ over $n$ network uses.
We obtain
\begin{align}
    nR 
    & = H(W) \nonumber \\
    & \stackrel{{\rm{(a)}}}{=} H \left (W \right ) - H \left (W|\mathcal{T}_{\mathcal{E}^-_{\mathsf{D}}}^{[n]} \right ) \nonumber \\
    & = I \left (W;\mathcal{T}_{\mathcal{E}^-_{\mathsf{D}}}^{[n]} \right ) \nonumber \\
    & \stackrel{{\rm{(b)}}}{\leq} I \left(W; \bigcup_{i=1}^{H_v} \mathcal{T}_{\mathcal{E}_i}^{[n]} \right) \nonumber \\
    & = I\left(W; \mathcal{T}_{\mathcal{Z}}^{[n]}\right) + I\left(W; \bigcup_{i=1}^{H_{v}} \mathcal{T}_{\mathcal{E}_{i} \setminus \mathcal{Z}}^{[n]} \;\middle|\; \mathcal{T}_{\mathcal{Z}}^{[n]}\right) \nonumber \\
    & \stackrel{{\rm{(c)}}}{\leq} \varepsilon + H\left( \bigcup_{i=1}^{H_{v}} \mathcal{T}_{\mathcal{E}_{i} \setminus \mathcal{Z}}^{[n]} \;\middle|\; \mathcal{T}_{\mathcal{Z}}^{[n]}\right)  \nonumber \\
    & \stackrel{{\rm{(d)}}}{\leq} \varepsilon + \sum_{i=1}^{H_v} H\left(  \mathcal{T}_{\mathcal{E}_{i} \setminus \mathcal{Z}_i}^{[n]} \;\middle|\; \mathcal{T}_{\mathcal{Z}_i}^{[n]}\right) \nonumber \\
    & \stackrel{{\rm{(e)}}}{\leq} \varepsilon + \sum_{i=1}^{H_v} \frac{h_i-K_i}{h_i} H\left(  \mathcal{T}_{\mathcal{E}_{i} }^{[n]}\right) \nonumber \\
    & \stackrel{{\rm{(f)}}}{\leq} \varepsilon + \sum_{i=1}^{H_v} \frac{h_i-K_i}{h_i} n,
    \label{eq:outer_proof3}
\end{align}
where the labeled (in)equalities follow from:
$\rm{(a)}$ the constraint for reliable decoding;
$\rm{(b)}$ the data processing inequality;
$\rm{(c)}$ the security constraint in~\eqref{eq:sec} and the fact that the entropy of a discrete random variable is non-negative;
$\rm{(d)}$ letting $\mathcal{Z}_i = \mathcal{E}_i\cap \mathcal{Z}$ and using the chain rule for the entropy and the fact that conditioning does not increase the entropy;
$\rm{(e)}$ applying~\cite[Lemma~1]{agarwal2018secure};
and $\rm{(f)}$ the 1-2-1 network constraint.

By dividing both sides of~\eqref{eq:outer_proof3} by $n$, we arrive at
\begin{equation}
R \leq \sum_{i=1}^{H_v} \frac{h_i-K_i}{h_i}.
\end{equation}
The above bound holds for any $\mathcal{Z}\subseteq \mathcal{E}$ such that $|\mathcal{Z}| = K$. Thus, we can find the tightest upper bound by minimizing it with respect to $K_i$'s, which yields
\begin{align}\label{eq:outer_proof4}
    R
    & \leq  \min_{\substack{K_i\in\mathbb{N}\cup\{0\},i\in[H_v]: \\ \sum_{i=1}^{H_v} K_i = K, \\ K_i \leq h_i,~i\in[H_v]}} \sum_{i=1}^{H_v} \frac{h_i - K_i}{h_i}.
\end{align}
Now, recall that the $h_i$'s are assumed (without loss of generality) to be sorted in descending order, i.e., $h_1 \geq h_2 \geq \ldots \geq h_{H_v}$.
This implies that a solution to~\eqref{eq:outer_proof4} would first fill $K_{H_v}$ with its maximum value, i.e., $K_{H_v} = \min \left \{ h_{H_v}, K\right \}$. Then, it will fill $K_{H_v-1}$ as $K_{H_v-1} = \min \left \{h_{H_v-1}, K - K_{H_v}  \right \}$ and so on until $K_1 = \min \left \{ h_1, K - \sum_{i=2}^{H_v} K_i\right \}$.
This concludes the proof of Theorem~\ref{thm:UB}.

\subsection{Proof of Theorem~\ref{thm:SecCap}}
\label{sec:ProofThmCap}
Without loss of generality, assume that $h_1 \geq h_2 \geq \ldots \geq h_{H_v}$.
If $M\geq H_v$, which implies $\widehat{M}=H_v$, it is not difficult to see that (see also Algorithm~\ref{alg:algo1}),
\begin{align}
    \mathsf{P}_{\widehat{M}} 
    & = \mathsf{P}_{H_v} 
    = h_{H_v}.
\end{align}
Then, the representation for $ G\setminus \mathcal{P}_{\widehat{M}}$ is given by $[h_1-h_{H_v}, \ldots, h_{H_v-1}-h_{H_v},0]$, which similarly gives
\begin{align}
    \mathsf{P}_{\widehat{M}-1}
    & = \mathsf{P}_{H_v-1}
    = h_{H_v-1} - h_{H_v}.
\end{align}
Iterating the above procedure up to $\mathcal{P}_1$, we obtain
\begin{align}\label{eq:matching_proof1}
    \mathsf{P}_{\ell}
    & = h_{\ell} - h_{\ell+1},~\forall \ell\in[H_v],
\end{align}
where we let $h_{H_v+1} = 0$.
Substituting~\eqref{eq:matching_proof1} into~\eqref{eq:Achievable} yields that for $c\in[0:H_v-1]$,
\begin{align}\label{eq:eq:matching_proof2}
    R(c)
    & = \left [ \frac{\sum_{\eta=0}^{c}(H_v-\eta)(h_{H_v-\eta} - h_{H_v-\eta+1})  - K}{\sum_{\eta=0}^{c} \left ( h_{H_v-\eta} - h_{H_v-\eta+1} \right )}   \right ]^+\nonumber \\
    & = \left [ H_v - c + \frac{\sum_{\eta=0}^{c-1}h_{H_v-\eta} - K }{h_{H_v-c}} \right ]^+.
\end{align}
Now, we pick $c\in[0:H_v-1]$ such that $\sum_{i=0}^{c-1} h_{H_v-i} < K \leq \sum_{i=0}^{c} h_{H_v-i}$. Note that such a $c$ always exists.
We define $\alpha_i = \min \left\{  h_i, K - \sum_{j=i+1}^{H_v} \alpha_j \right\},~ i\in[H_v]$, that is,
\begin{align}
\label{eq:alphas}
    \alpha_i 
    & = \begin{cases}
        0 & \text{ if } i<H_v-c-1, \\
        K - \sum_{j=i+1}^{H_v} h_j & \text{ if } i = H_v-c, \\
        h_i & \text{ if } i > H_v-c.
    \end{cases}
\end{align}
The $\alpha_i$'s in~\eqref{eq:alphas} imply that
\begin{align}\label{eq:matching_proof3}
    \frac{K - \sum_{\eta=0}^{c-1}h_{H_v-\eta} }{h_{H_v-c}} + c
    & =  \frac{K - \sum_{\eta=0}^{c-1}h_{H_v-\eta} }{h_{H_v-c}} + \!\!\!\!\sum_{i=H_v-c+1}^{H_v} \frac{h_{i}}{h_{i}} \nonumber \\
    & = \sum_{i=1}^{H_v} \frac{\alpha_i}{h_i}.
\end{align}
Substituting~\eqref{eq:matching_proof3} into~\eqref{eq:eq:matching_proof2}, we obtain
\begin{align}
    R(c)
    & = \left [ H_v - \sum_{i=1}^{H_v} \frac{\alpha_i}{h_i} \right ]^+ = \sum_{i=1}^{H_v} \left( 1 - \frac{\alpha_i}{h_i} \right),
\end{align}
where
\begin{align}
    \alpha_i 
    & = \min \left\{  h_i, K - \sum_{j=i+1}^{H_v} \alpha_j \right\},~ i\in[H_v],
\end{align}
which is the upper bound in~\eqref{eq:UBThm}.
This concludes the proof of Theorem~\ref{thm:SecCap}.

\section*{Acknowledgment}
This research was supported in part by the U.S. National Science Foundation under Grant CCF-2045237.

%%%%%%
%% To balance the columns at the last page of the paper use this
%% command:
%%
%\enlargethispage{-1.2cm} 
%%
%% If the balancing should occur in the middle of the references, use
%% the following trigger:
%%

\appendices

\section{Proof of Lemma~\ref{lem:t}} \label{app:proof_t}

For $\ell=1$, $\mathcal{P}_{1}$ is simply the set of edge disjoint paths in $G$ and hence, $\mathsf{P}_1 \left (h_{[H_v]} \right ) =\sum_{i=1}^{H_v} h_{i}$. 
We now prove Lemma~\ref{lem:t} for $\ell \in [2:H_{v}]$ for two cases, separately. 

\noindent {\bf Case 1: $h_1 \geq \mathsf{P}_{\ell-1} \left (h_{[2:H_v]} \right )$.}
Note that $\mathcal{P}_{\ell-1} \left (h_{[2:H_v]}\right )$ is the set of groups of $\ell-1$ vertex disjoint paths selected from $h_{[2:H_v]}$ in $G$.
Now, since $h_1 \geq \mathsf{P}_{\ell-1} \left (h_{[2:H_v]} \right )$, we can construct $\mathcal{P}_{\ell} \left (h_{[H_v]} \right )$ as follows. We select $\mathsf{P}_{\ell-1} \left (h_{[2:H_v]} \right )$ different paths in $h_1$ and assign each of them to one of the groups of paths in $\mathcal{P}_{\ell-1} \left (h_{[2:H_v]}\right )$. The resulting $\mathcal{P}_{\ell} \left (h_{[H_v]} \right )$ will consist of $\mathsf{P}_{\ell-1} \left (h_{[2:H_v]} \right )$ groups of $\ell$ vertex disjoint paths selected from $h_{[H_v]}$ in $G$. Thus, $\mathsf{P}_{\ell} \left (h_{[H_v]} \right ) =\mathsf{P}_{\ell-1} \left (h_{[2:H_v]} \right )$.

\noindent {\bf Case 2: $h_{1} < \mathsf{P}_{\ell-1} \left (h_{[2:H_v]} \right )$.}
We propose a simple algorithm to construct $\mathcal{P}_{\ell} \left (h_{[H_v]} \right )$.
This algorithm runs in $B \geq 1$ rounds, where each round $b \in [B]$ consists of two phases.
In particular, we consider $H_v$ bins with ${h}_i^{(b)},~i\in[H_v]$ elements at the $b$th round of the algorithm. 
We initialize $\mathbf{h}^{(0)} = [h_1, h_2, \ldots , h_{H_v}]$.
The algorithms runs as follows,
\begin{itemize}
\item Phase~I of round $b \in [B]$: We remove one element from each of the first $\ell$ elements of ${\mathbf{h}}^{(b-1)}$ and construct $\tilde{\mathbf{h}}^{(b)}$, that is,
\begin{equation}
\tilde{\mathbf{h}}^{(b)} = \left \{
\begin{array}{ll}
h^{(b-1)}_i -1 & i \in [\ell],
\\
h^{(b-1)}_i &  i \in [\ell+1:H_v].
\end{array}
\right .
\end{equation}
If there is a number of non-empty bins less than $\ell$, we stop the algorithm.
\item Phase~II of round $b \in [B]$. We sort $\tilde{\mathbf{h}}^{(b)}$ in descending order and let $\mathbf{h}^{(b)}$ be the corresponding vector, i.e., ${h}^{(b)}_1 \geq {h}^{(b)}_2 \geq \ldots \geq {h}^{(b)}_{H_v}$.
\end{itemize}
The above algorithm terminates after $B$ rounds, where $B$ is such that $\mathbf{h}^{(B)}$ has at most $\ell-1$ elements different from zero, i.e., we have ${h}_i^{(B)} =0$ for at least all $i \in [\ell:H_v]$.
It is not difficult to see that $\mathsf{P}_{\ell} \left (h_{[H_v]} \right )$
%$\mathcal{P}_{\ell} \left (h_{[H_v]} \right )$ 
is equal to (at least) the number of rounds $B$ of the simple algorithm proposed above, i.e., $\mathsf{P}_{\ell} \left (h_{[H_v]} \right ) = B$. Hence, we need to find the value of $B$.
Note that
\begin{equation}
\label{eq:PerfAlgo}
t_{\ell}^{\mathsf{algo}}\left (h_1,h_2,\ldots,h_{H_v} \right ) = B,
\end{equation}
where $t_{\ell}^{\mathsf{algo}}\left (h_1,h_2,\ldots,h_{H_v} \right )$ denotes the number of rounds of the above algorithm.

We next prove the following property of the above described algorithm.
\begin{prop}
    \label{prop:Induct}
Assume that there exists $q \in [0:B-1]$ such that, for an integer $x \geq 2$,
\begin{equation}
{h}_1^{(B-q)} = q+x \quad  \text{and} \quad {h}_{\ell}^{(B-q)} \in [0:q].
\end{equation}
Then, it holds that
\begin{equation}
{h}_1^{(B-(q+1))} = {h}_1^{(B-q)}+1 \quad  \text{and} \quad {h}_{\ell}^{(B-(q+1))} \in [1:q+1].
\end{equation}
\end{prop}
\begin{IEEEproof}
We start by noting that the condition ${h}_{\ell}^{(B-q)} \in [0:q]$, together with the fact that at each round $b \in [1:B]$ the vector $\mathbf{h}^{(b)}$ is sorted in descending order, readily implies that ${h}_{\ell}^{(B-(q+1))} \in [1:q+1]$.

We now observe that the fact that ${h}_1^{(B-q)} = q+x$ implies that we have either one of the two following cases:
\begin{itemize}
\item Case~1: ${h}_1^{(B-(q+1))} = q+x +1$, 
\item Case~2: ${h}_1^{(B-(q+1))} = q+x$ and ${h}_{\ell}^{(B-(q+1))} = q+x$.
\end{itemize}
However, we have shown above that ${h}_{\ell}^{(B-(q+1))} \in [1:q+1]$, which implies that Case~2 cannot happen since it is assumed that $x \geq 2$. Moreover, Case~1 reduces to
\begin{equation}
{h}_1^{(B-(q+1))} = q+x+1 = {h}_1^{(B-q)} +1,
\end{equation}
and
\begin{equation}
{h}_{\ell}^{(B-(q+1))} \in [1:q+1].
\end{equation}
This concludes the proof of Property~\ref{prop:Induct}.
\end{IEEEproof}

We now leverage Property~\ref{prop:Induct} to prove the following additional property of the above algorithm. We denote with $T$ the number of elements of $\mathbf{h}^{(B)}$ 
%$\mathbf{h}^{(K)}$ 
that are different from zero.
\begin{prop}
    \label{prop:Simplification}
Under the condition  $h_1 < t_{\ell-1}^{\mathsf{algo}}\left (h_2,\ldots,h_{H_v} \right )$ with $h_1 \triangleq {h}_1^{(0)}$,  we have that ${h}_i^{(B)} = 1$ for all $i \in [T]$.
\end{prop}

\begin{IEEEproof}
The proof is by contradiction. In particular, we assume that ${h}_1^{(B)} >1$ and we show that it leads to a contradiction. We let ${h}_1^{(B)} =x$, where $x \geq 2$. We note that, since $T < \ell$, we have that ${h}_\ell^{(B)} =0$.
Therefore, $q=0$ satisfies Property~\ref{prop:Induct} and it can be used as the base case of a proof by induction. By inductively using Property~\ref{prop:Induct}, we arrive at
\begin{equation}
h_1 \triangleq {h}_1^{(0)} = B+ x.
\end{equation}
This, together with the fact that ${h}_1^{(B)} =x$, implies that we remove an element from $h_1$ at each of the $B$ steps of the algorithm.
Therefore,
\begin{equation}
t_{\ell-1}^{\mathsf{algo}}\left (h_2,\ldots,h_{H_v} \right ) = B.
\end{equation}
The above leads to
\begin{equation}
t_{\ell-1}^{\mathsf{algo}}\left (h_2,\ldots,h_{H_v} \right ) = B < B+x = h_1,
\end{equation}
which leads to a contradiction since we are considering $h_1 < t_{\ell-1}^{\mathsf{algo}}\left (h_2,\ldots,h_{H_v} \right )$.
This concludes the proof of Property~\ref{prop:Simplification}.
\end{IEEEproof}

We now leverage Property~\ref{prop:Simplification} to conclude the proof of Lemma~\ref{lem:t}.
We note that
\begin{equation}
B = \frac{\sum_{i=1}^{H_v} h_i-T}{\ell}.
\end{equation}
The above follows since:
(i) before the algorithm described above is run, we have a total number of elements equal to $\sum_{i=1}^{H_v} h_i$;
(ii) Property~\ref{prop:Simplification} shows that the total number of elements remaining after the $B$ rounds of the algorithm is $T < \ell$;
and
(iii) at each of the $B$ rounds of the algorithm, we remove $\ell$ elements.

Now, by using~\eqref{eq:PerfAlgo}, we have that
\begin{equation}
t_{\ell}^{\mathsf{algo}}\left (h_1,h_2,\ldots,h_{H_v} \right ) = \frac{\sum_{i=1}^{H_v} h_i-T}{\ell},
\end{equation}
which implies
\begin{equation}
t_{\ell}^{\mathsf{algo}}\left (h_1,h_2,\ldots,h_{H_v} \right ) + \frac{T}{\ell} = \frac{\sum_{i=1}^{H_v} h_i}{\ell},
\end{equation}
and hence,
\begin{equation}
\left \lfloor t_{\ell}^{\mathsf{algo}}\left (h_1,h_2,\ldots,h_{H_v} \right ) + \frac{T}{\ell} \right \rfloor = \left \lfloor \frac{\sum_{i=1}^{H_v} h_i}{\ell} \right \rfloor.
\end{equation}
Finally, since $t_{\ell}^{\mathsf{algo}}\left (h_1,h_2,\ldots,h_{H_v} \right )$ is an integer and $T < \ell$, the above is equivalent to
\begin{equation}
t_{\ell}^{\mathsf{algo}}\left (h_1,h_2,\ldots,h_{H_v} \right ) =  \left \lfloor \frac{\sum_{i=1}^{H_v} h_i}{\ell} \right \rfloor.
\end{equation}
The proof of Lemma~\ref{lem:t} is concluded by observing that
\begin{equation}
    t_{\ell}^{\mathsf{algo}}\left (h_1,h_2,\ldots,h_{H_v} \right ) 
    = \mathsf{P}_{\ell} \left (h_{[H_v]} \right ).
\end{equation}

\IEEEtriggeratref{15}
%%
%% which triggers a \newpage (i.e., new column) just before the given
%% reference number. Note that you need to adapt this if you modify
%% the paper.  The "triggered" command can be changed if desired:
%%
%\IEEEtriggercmd{\enlargethispage{-20cm}}
%%
%%%%%%

\bibliographystyle{IEEEtran}
\bibliography{myref}

\begin{thebibliography}{10}
\providecommand{\url}[1]{#1}
\csname url@samestyle\endcsname
\providecommand{\newblock}{\relax}
\providecommand{\bibinfo}[2]{#2}
\providecommand{\BIBentrySTDinterwordspacing}{\spaceskip=0pt\relax}
\providecommand{\BIBentryALTinterwordstretchfactor}{4}
\providecommand{\BIBentryALTinterwordspacing}{\spaceskip=\fontdimen2\font plus
\BIBentryALTinterwordstretchfactor\fontdimen3\font minus
  \fontdimen4\font\relax}
\providecommand{\BIBforeignlanguage}[2]{{%
\expandafter\ifx\csname l@#1\endcsname\relax
\typeout{** WARNING: IEEEtran.bst: No hyphenation pattern has been}%
\typeout{** loaded for the language `#1'. Using the pattern for}%
\typeout{** the default language instead.}%
\else
\language=\csname l@#1\endcsname
\fi
#2}}
\providecommand{\BIBdecl}{\relax}
\BIBdecl

\bibitem{Wang2018}
X.~Wang, L.~Kong, F.~Kong, F.~Qiu, M.~Xia, S.~Arnon, and G.~Chen, ``Millimeter
  {W}ave {C}ommunication: A {C}omprehensive {S}urvey,'' \emph{IEEE
  Communications Surveys \& Tutorials}, vol.~20, no.~3, pp. 1616--1653, 2018.

\bibitem{Uwaechia2020}
A.~N. Uwaechia and N.~M. Mahyuddin, ``A {C}omprehensive {S}urvey on
  {M}illimeter {W}ave {C}ommunications for {F}ifth-{G}eneration {W}ireless
  {N}etworks: {F}easibility and {C}hallenges,'' \emph{IEEE Access}, vol.~8, pp.
  62\,367--62\,414, 2020.

\bibitem{ezzeldin2020gaussian}
Y.~H. Ezzeldin, M.~Cardone, C.~Fragouli, and G.~Caire, ``Gaussian 1-2-1
  {N}etworks: {C}apacity {R}esults for mm{W}ave {C}ommunications,'' \emph{IEEE
  Transactions on Information Theory}, vol.~67, no.~2, pp. 961--990, 2020.

\bibitem{agarwal2018secure}
G.~K. Agarwal, Y.~H. Ezzeldin, M.~Cardone, and C.~Fragouli, ``Secure
  {C}ommunication over 1-2-1 {N}etworks,'' in \emph{2018 IEEE International
  Symposium on Information Theory (ISIT)}, 2018, pp. 196--200.

\bibitem{EzzeldinSecurity}
Y.~H. Ezzeldin, M.~Cardone, and C.~Fragouli, ``Multilevel {S}ecrecy over 1-2-1
  {N}etworks,'' in \emph{2020 IEEE Information Theory Workshop (ITW)}, 2020,
  pp. 1--5.

\bibitem{Mukherjee2011}
A.~Mukherjee and A.~L. Swindlehurst, ``Robust {B}eamforming for {S}ecurity in
  {MIMO} {W}iretap {C}hannels with {I}mperfect {CSI},'' \emph{IEEE Transactions
  on Signal Processing}, vol.~59, no.~1, pp. 351--361, 2011.

\bibitem{safaka2016creating}
I.~Safaka, L.~Czap, K.~Argyraki, and C.~Fragouli, ``Creating {S}ecrets {O}ut
  {O}f {P}acket {E}rasures,'' \emph{IEEE Transactions on Information Forensics
  and Security}, vol.~11, no.~6, pp. 1177--1191, 2016.

\bibitem{cai2002secure}
N.~Cai and R.~Yeung, ``Secure {N}etwork {C}oding,'' in \emph{2002 IEEE
  International Symposium on Information Theory,}, 2002, pp. 323--.

\bibitem{khaleghi2009subspace}
A.~Khaleghi, D.~Silva, and F.~R. Kschischang, ``Subspace {C}odes,'' in
  \emph{IMA International Conference on Cryptography and Coding}.\hskip 1em
  plus 0.5em minus 0.4em\relax Springer, 2009, pp. 1--21.

\bibitem{jaggi2007resilient}
S.~Jaggi, M.~Langberg, S.~Katti, T.~Ho, D.~Katabi, and M.~M{\'e}dard,
  ``Resilient {N}etwork {C}oding in the {P}resence of {B}yzantine
  {A}dversaries,'' in \emph{2007-26th IEEE {I}nternational {C}onference on
  {C}omputer {C}ommunications (INFOCOM)}.\hskip 1em plus 0.5em minus
  0.4em\relax IEEE, 2007, pp. 616--624.

\bibitem{jaggi2005polynomial}
S.~Jaggi, P.~Sanders, P.~A. Chou, M.~Effros, S.~Egner, K.~Jain, and L.~M.
  Tolhuizen, ``Polynomial {T}ime {A}lgorithms for {M}ulticast {N}etwork {C}ode
  {C}onstruction,'' \emph{IEEE Transactions on Information Theory}, vol.~51,
  no.~6, pp. 1973--1982, 2005.

\bibitem{agarwal2017secure}
G.~K. Agarwal, M.~Cardone, and C.~Fragouli, ``Secure {N}etwork {C}oding for
  {M}ultiple {U}nicast: {O}n the {C}ase of {S}ingle {S}ource,'' in
  \emph{Information Theoretic Security: 10th International Conference
  (ICITS)}.\hskip 1em plus 0.5em minus 0.4em\relax Springer, 2017, pp.
  188--207.

\bibitem{cui2012secure}
T.~Cui, T.~Ho, and J.~Kliewer, ``On {S}ecure {N}etwork {C}oding with
  {N}onuniform or {R}estricted {W}iretap {S}ets,'' \emph{IEEE Transactions on
  Information Theory}, vol.~59, no.~1, pp. 166--176, 2012.

\bibitem{cai2007security}
N.~Cai and R.~W. Yeung, ``A {S}ecurity {C}ondition for {M}ulti-{S}ource
  {L}inear {N}etwork {C}oding,'' in \emph{2007 IEEE International Symposium on
  Information Theory (ISIT)}, 2007, pp. 561--565.

\bibitem{ngai2011network}
C.-K. Ngai, R.~W. Yeung, and Z.~Zhang, ``Network {G}eneralized {H}amming
  {W}eight,'' \emph{IEEE Transactions on Information Theory}, vol.~57, no.~2,
  pp. 1136--1143, 2011.

\bibitem{wei1991generalized}
V.~K. Wei, ``Generalized {H}amming {W}eights for {L}inear {C}odes,'' \emph{IEEE
  Transactions on Information Theory}, vol.~37, no.~5, pp. 1412--1418, 1991.

\bibitem{koetter2004network}
R.~Koetter, M.~Effros, T.~Ho, and M.~M{\'e}dard, ``Network codes as codes on
  graphs,'' in \emph{2004 Annual Conference on Information Sciences and Systems
  (CISS)}, 2004.

\bibitem{gadouleau2011graph}
M.~Gadouleau and S.~Riis, ``Graph-{T}heoretical {C}onstructions for {G}raph
  {E}ntropy and {N}etwork {C}oding {B}ased {C}ommunications,'' \emph{IEEE
  Transactions on Information Theory}, vol.~57, no.~10, pp. 6703--6717, 2011.

\bibitem{ho2004byzantine}
T.~Ho, B.~Leong, R.~Koetter, M.~Medard, M.~Effros, and D.~R. Karger,
  ``Byzantine {M}odification {D}etection in {M}ulticast {N}etworks {W}ith
  {R}andom {N}etwork {C}oding,'' \emph{IEEE Transactions on Information
  Theory}, vol.~54, no.~6, pp. 2798--2803, 2008.

\bibitem{papadopoulos2015lp}
A.~Papadopoulos, L.~Czap, and C.~Fragouli, ``L{P} formulations for secrecy over
  erasure networks with feedback,'' in \emph{2015 IEEE International Symposium
  on Information Theory (ISIT)}, 2015, pp. 954--958.

\bibitem{czap2014triangle}
L.~Czap, V.~M. Prabhakaran, S.~Diggavi, and C.~Fragouli, ``Triangle {N}etwork
  {S}ecrecy,'' in \emph{2014 IEEE International Symposium on Information
  Theory}, 2014, pp. 781--785.

\end{thebibliography}

\end{document}